\documentclass[intlimits,twoside,a4paper]{article}

\usepackage{amsmath,amssymb}
\usepackage{graphicx}

\usepackage[T2A]{fontenc}
\usepackage[cp1251]{inputenc}

\usepackage[eqsecnum]{cmpj3}

\issue{2019}{22}{4}{43602}
\doinumber{10.5488/CMP.22.43602}

\title{Detection of collective optic excitations in molten NaI}%
\author[S. Hosokawa \textsl{et al.}]{S. Hosokawa\refaddr{label1,label2}\thanks{Electronic mail address: shhosokawa@kumamoto-u.ac.jp}\,, M. Inui\refaddr{label3}, T. Bryk\refaddr{label4}, I. Mryglod\refaddr{label4}, W.-C. Pilgrim\refaddr{label2}, Y. Kajihara\refaddr{label3}, K.~Matsuda\refaddr{label5}, Y. Ohmasa\refaddr{label6}, S. Tsutsui\refaddr{label7}}
\addresses{
\addr{label1} Department of Physics, Kumamoto University, Kumamoto 860-8555, Japan
\addr{label2} Department of Chemistry, Physical Chemistry, Philipps University of Marburg, 35032 Marburg, Germany
\addr{label3} Graduate School of Integrated Arts and Sciences, Hiroshima University, Higashi-Hiroshima 739-8521, Japan
\addr{label4} Institute for Condensed Matter Physics of the National Academy of Sciences of Ukraine, 79011 Lviv, Ukraine
\addr{label5} Department of Physics, Kyoto University, Kyoto 606-8502, Japan
\addr{label6} Neutron Science Laboratory, Institute for Solid State Physics, The University of Tokyo, Kashiwa 277-8581, Japan
\addr{label7} Japan Synchrotron Radiation Research Institute (JASRI), Sayo 679-5198, Japan
}

\date{Received June 24, 2019}

\begin{document}

\maketitle

\begin{abstract}
High-resolution inelastic x-ray scattering measurements were carried out on molten NaI near the melting point at 680$^\circ$C at SPring-8. Small and damped indications of longitudinal optic excitation modes were observed on the tails of the longitudinal acoustic modes at small momentum transfers, $Q\sim5$ nm$^{-1}$. The measured spectra are in good agreement, in both frequency and linewidth, with {\it ab initio} molecular dynamics (MD) simulations but not classical MD simulations. The observation of these modes at small $Q$ and a good agreement with the simulation permits clear identification of these as collective optic modes with well defined phasing between different ionic motions.  

\pacs 61.10.Eq, 61.20.Lc, 78.30.Cp
\end{abstract}

The existence of optic excitation modes in molten salts was predicted by Hansen and McDonald~\cite{Hansen} more than four decades ago by a pioneering molecular dynamics (MD) simulation, which revealed the existence of propagating short wavelength charge fluctuations as in the solid state. Further MD simulations with more advanced potentials confirmed this prediction by including mass effects and ion polarization \cite{Adams,Dixon}. These computer studies motivated several groups to investigate ion dynamics in molten salts by inelastic neutron scattering \cite{Price,McGreevy}.  However, the experiments were subject to the kinematic constraints of the energy ($\hbar\omega$)$-$momentum ($Q$) transfer relation of the neutrons. Thus, the neutron data were measured at $Q>10$ nm$^{-1}$, where optic modes may be highly damped and/or buried beneath experimental backgrounds. Inelastic x-ray scattering (IXS) has nearly no such kinematic constraints and tends to have reduced backgrounds. 

In a two component molten salt, the dynamic structure factor, $S(Q,\omega)$, is written as a combination of three scattering functions \cite{HansenBook},
\begin{eqnarray}
\label{Partials}
S(Q,\omega)&\propto&(f_++f_-)^2S_{\rm NN}(Q,\omega)
+(f_+^2-f_-^2)S_{\rm NZ}(Q,\omega)+(f_+-f_-)^2S_{\rm ZZ}(Q,\omega),
\end{eqnarray}
where $f_+$ and $f_-$ are atomic form factors of cation and anion atoms, respectively. $S_{\rm NN}$ is the density fluctuations (acoustic modes), $S_{\rm ZZ}$ is determined by charge fluctuations (optic modes), and $S_{\rm NZ}$ is the cross term. As can be inferred from this equation, a large difference in the $f$ values gives a large weight to the charge fluctuations. Thus, NaI seems to be one of the most favorable molten salts to detect the optic excitation modes.

The observation of an optic mode at small $Q$ is important, as an analysis by Bryk and Mryglod \cite{BrykNaCl} using an approach of generalized collective modes (GCM) shows a significant change in mode character as $Q$ increases.  In the small $Q$ region, $\lesssim 10$ nm$^{-1}$,  the low-$\omega$ acoustic mode reflects a longitudinal propagating pressure wave and the high-$\omega$ optic mode reflects the out-of-phase charge current fluctuations.  In the $Q$ region beyond  10 nm$^{-1}$, however, the low- or high-$\omega$ branches reflect primarily the dynamics of the heavy or light sub-system, respectively.  Thus, observation of modes at low $Q$ demonstrates the presence of a true out-of-phase motion, which is characteristic of a longitudinal optic (LO) mode in a solid.

In this letter, we report measurements of the LO collective excitation modes in molten NaI obtained from a precise IXS experiment with  good statistics and a low background. The LO modes were found in the low $Q$ range, $\sim5$ nm$^{-1}$, as small and damped shoulders on the tail of the longitudinal acoustic (LA) excitation modes, in excellent agreement with the results of an {\it ab initio} MD simulation. The present results are the first experimental proof for these theoretically predicted excitations at small $Q$, and the first unambiguous observation of optic modes in a molten salt. 

The $S(Q,\omega)$ spectra of molten NaI were measured at 680$^\circ$C at BL35XU/SPring-8 \cite{Baron} using the Si(11 11 11) reflection with an incident x-ray energy of 21.747 keV. In order to obtain  better statistics, the measurements were performed at one goniometer angles, corresponding to $Q$ values between 1.3 and 5.5 nm$^{-1}$, and more than twenty scans (almost two days) were performed for investigating the small LO modes. The energy resolution was about 1.4 meV, and the $Q$ resolution was about $\pm0.4$ nm$^{-1}$. The $\omega$ range was $\pm40$ meV due to the high excitation energies of the LO modes. This high-intensity low-background IXS spectrometer \cite{BaronInstrum} allows one to observe small non-hydrodynamic excitations, such as transverse acoustic (TA) excitations in a simple liquid metal \cite{HosokawaPRL,11,12}.

A previous IXS experiment on molten NaI \cite{Demmel} was performed with poorer statistics and showed small indications of the LO modes. Since the tiny signals of LO modes are buried in the background produced by the sapphire sample container, mainly LA and TA excitations of sapphire, this was not a clear evidence of the observation of LO modes in a molten salt and we regard the present experiment with better statistics and lower background as correct. 

The sample was contained in a cell made of single crystal sapphire \cite{Tamura,HosokawaCMP}. In order to suppress the background from the sapphire container, the x-ray windows were polished down to 0.15 mm. It was placed in an internally heated vessel \cite{HosokawaCMP,HosokawaRSI} with x-ray windows made of thin Si single crystals capable of covering scattering angles between 0$^\circ$ and 25$^\circ$. A pressure of 1.5 bar of high-purity-grade He gas was applied to stabilize the sample. The temperature of 680$^\circ$C was achieved using a Mo resistance wire of 0.5 mm in diameter, and monitored with two W-Re thermocouples.

Theoretical time-correlation functions for molten NaI were obtained from an {\it ab initio} MD simulation. A system of 150 particles was simulated in a constant volume and temperature (680$^\circ$C) ensemble. A detailed calculation procedure is given in \cite{Bryk}.

\begin{figure}[!t]
\centering
\includegraphics[width=.95\textwidth]{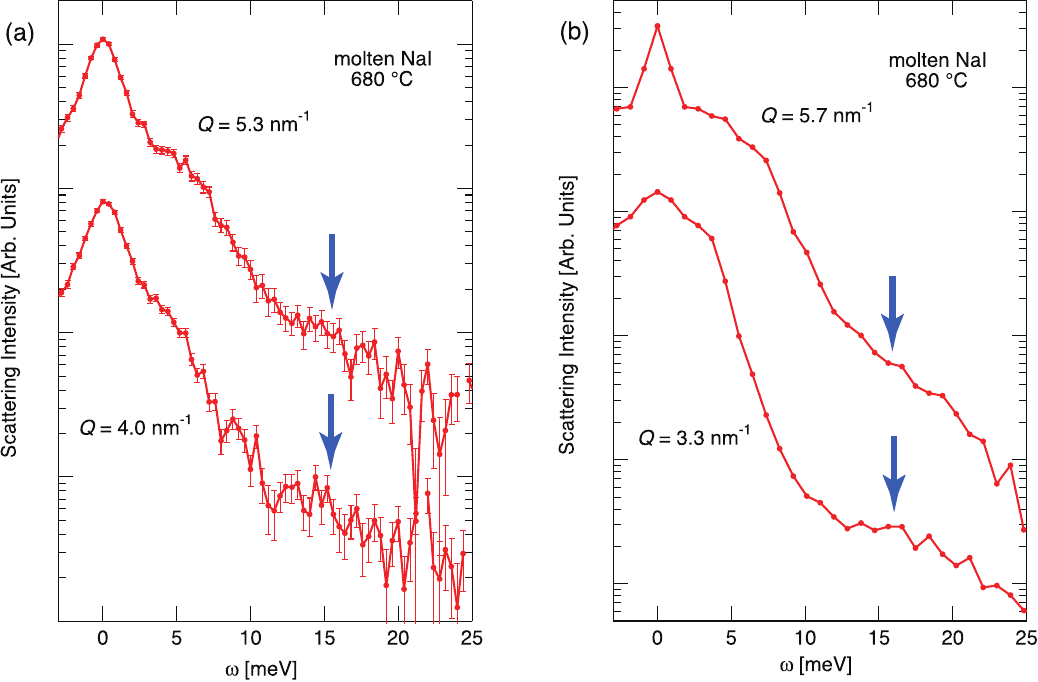}
\caption{\label{Results} (Colour online) Logarithmic plots of IXS spectra of molten NaI at 680$^\circ$C (a) measured at $Q=4.0$ and 5.3 nm$^{-1}$ and (b) calculated by an {\it ab initio} MD simulation at $Q=3.3$ and 5.7 nm$^{-1}$.}
\end{figure}

Figure~\ref{Results}~(a) shows logarithmic plots of the scattering intensities measured at $Q=4.0$ and 5.3 nm$^{-1}$. The LA excitations, as is typical in most liquids, are easily visible as shoulders at $\omega\sim4{-}6$ meV. The shape of the scattering intensity on the tails of these shoulders clearly indicates additional intensity at about 15--17 meV as indicated by the arrows. The intensity is about one hundredth of the central quasielastic peaks, and about one twentieth of the LA excitations. 

In the lower $Q$ range, the observed intensity is hampered by the LA- and TA modes from the sapphire container. Unfortunately, at $Q=1.5$ nm$^{-1}$ where the sapphire LA modes do not interfere with the LO modes of molten NaI,  the high frequency excitations are too small to be detected, reflecting the $Q^2$ vanishing feature of the high frequency LO modes in the low $Q$ range \cite{BrykNaCl}.

\begin{figure}[!t]
\centering
\vspace{3mm}
\includegraphics[width=.45\textwidth]{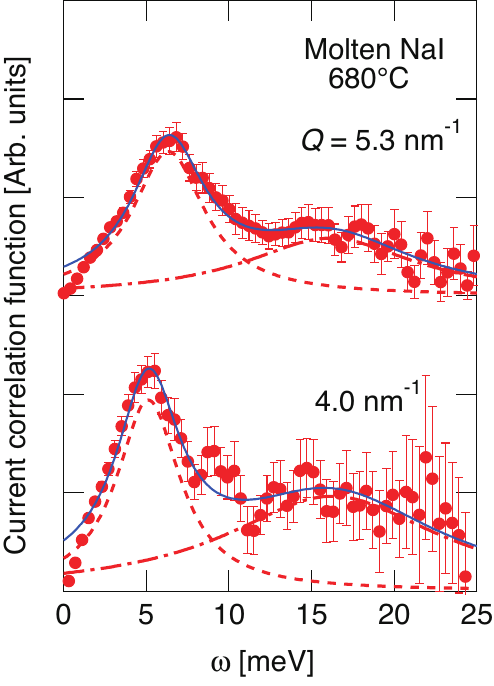}
\caption{\label{CCF} (Colour online) Current correlation functions of molten NaI obtained from the experimental data shown in figure~\ref{Results}~(a). The chain and dashed curves indicate the LO and LA collective excitation contributions, respectively.}
\end{figure}

This experimentally observed extra intensity is also observed in the {\it ab initio} MD simulations. In order to compare with the theoretical data, calculated x-ray scattering intensity at $Q=3.3$ and 5.7~nm$^{-1}$ are given in figure~\ref{Results}~(b). The spectral features of the experimental and theoretical results are very similar. In both cases, broad shoulders appear at about $15{-}17$ meV as indicated by the arrows. In the MD simulation, this extra intensity is clearly identified as a high frequency branch of the longitudinal excitations indicating LO-like collective modes. The present IXS result is the first reliable experimental proof for the existence of LO excitation modes in a simple molten salt.

For the analysis of the data, we used the current correlation function, $J(Q,\omega)=({\omega^2}/{Q^2})S(Q,\omega)$. This representation of the scattering data suppresses the central quasielastic intensity, and emphasizes the excitations in the high $\omega$ range. Figure \ref{CCF} shows $J(Q,\omega)$ at $Q=4.0$ and 5.3 nm$^{-1}$. As clearly seen in the figure, a sharp peak at about $5{-}6$ meV and a broad peak at about $15{-}17$ meV appear in each spectrum. 
The low $\omega$ peaks correspond to the LA collective excitations, and the high $\omega$ peaks correspond to the LO ones. Two Lorentzians were used for the analysis of the spectra giving the peak energies and linewidths. In the figure, the solid curves are the best fits, and the dashed and chain curves represent the LA and LO collective excitation contributions, respectively. The LA collective excitations show a dispersion, i.e., their energies change with $Q$. However, the  dispersion of LO excitations is very weak.

\begin{figure}[!t]
\centering
\includegraphics[width=.49\textwidth]{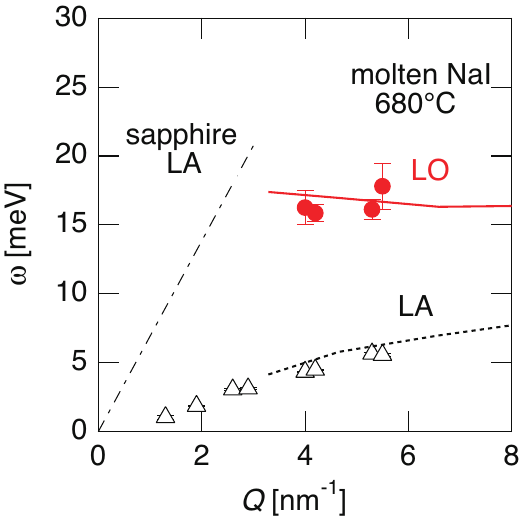}
\caption{\label{Dispersion} (Colour online) Dispersion relation of the collective excitations in molten NaI. Circles and triangles indicate the experimental data for the LO and LA modes, respectively, and the solid and dashed curves show the theoretical ones \cite{Bryk}. See the text for details.}
\end{figure}

Circles in figure~\ref{Dispersion} show the dispersion relation of the LO collective excitations together with that of the LA excitation modes indicated by triangles which were obtained from the usual damped harmonic oscillator (DHO) fits \cite{DHO} to avoid the effect of the quasielastic contributions. Solid and dashed curves in the figure are the theoretical results, where the former is the peak positions of the theoretical $J(Q,\omega)$ functions for the LO modes, and the latter is the theoretical dispersion for the LA excitation modes, i.e., the imaginary part of the complex collective eigenvalues, $\Im[z_\alpha(Q)]$. The excitation energies of both the LO and LA collective excitation modes obtained from theory are in excellent agreement with the present experimental results. The chain line indicates the dispersion relation of the LA collective excitations from the sapphire container, which prevents an unambiguous indication of the LO modes in the low $Q$ range.

\begin{figure}[!t]
\centering
\includegraphics[width=.5\textwidth]{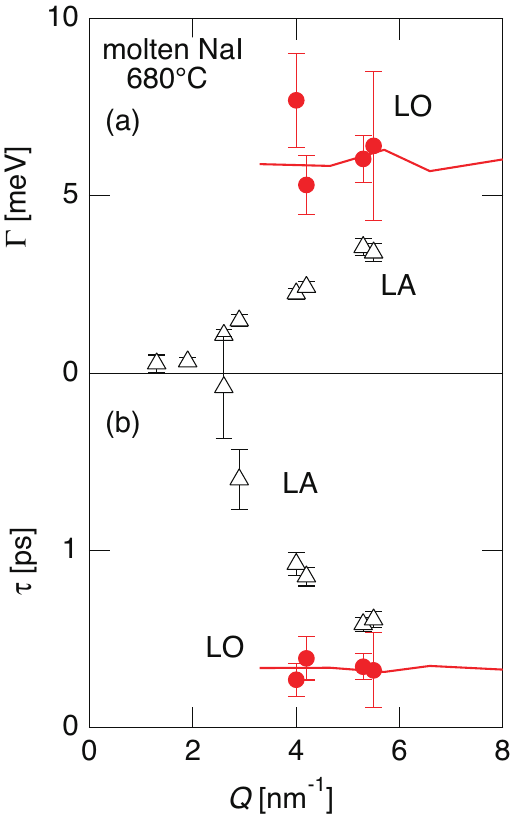}
\caption{\label{Lifetime} (Colour online) (a) The width $\Gamma$ and (b) lifetime $\tau$ of the LO (open circles) and LA (closed triangles) excitations, and the solid curves the theoretical ones for the LO modes. See the text for details.}
\end{figure}

Circles in figure~\ref{Lifetime}~(a) show the values of half width at half maximum (HWHM), $\Gamma$, of the LO collective excitations, where the widths of the experimental resolution functions were taken into account. The widths of the LA collective excitation modes are also indicated by triangles in the figure, which are obtained from the DHO fits, and very close to the HWHM values. In the same figure, theoretical damping values, i.e., the widths of the theoretical $J(Q,\omega)$ functions for the LO modes are given by a solid curve. The theoretical $\Gamma$ values for the LO modes again coincide well with the experimental results. 

From the spectral widths of the collective excitations, the corresponding lifetimes, $\tau(Q)$, can be estimated to be proportional to the inverse of the widths, i.e., $\tau(Q)=h/(2\hbar\Gamma)= \piup/\Gamma$ by taking the Heisenberg's uncertainty principle into account, where $h$ is the Planck constant and $\hbar=h/2\piup$. Circles and triangles in figure~\ref{Lifetime}~(b) show the experimentally determined lifetimes of the LO and LA collective excitations, respectively, and the solid curve represents the theoretical ones for the LO modes. The $\tau$ values of the LO collective modes are about 0.3 ps, and almost independent of $Q$. On the other hand, the $\tau$ values of the LA collective modes are larger than the LO values by a factor of 2--3, and rapidly increase with decreasing $Q$.

The existence of high-$\omega$ excitation modes was reported from several inelastic neutron scattering investigations on liquid binary mixtures, such as liquid Li$_4$Pb \cite{Alvarez}, Na$_x$Sn$_{1-x}$ \cite{Jahn}, or DF \cite{Bermejo}. Due to the kinematic constraints, however, such high-$\omega$ modes were found only in the high $Q$ range beyond $\sim10$~nm$^{-1}$. In this $Q$ region, the high- and low-$\omega$ branches of binary liquids are interpreted by reflecting mainly partial dynamics of light and heavy elements, respectively, as mentioned in the introductory. Contrary to this, the two branches of collective excitations invested here have different origins; a high-$\omega$ branch from non-hydrodynamic optic excitations and a low-$\omega$ one from hydrodynamic acoustic modes. The optic high-$\omega$ branch in molten NaI originates from motions of differently charged species with opposite phases, which is confirmed by a prominent excitation peak in $S_{\rm ZZ}(Q,\omega)$ calculated in a classical~\cite{Alcaraz} MD simulation. The existence of optic modes was also discussed for the IXS results of water as the low-$\omega$ branch in relation to the fast sound \cite{Sette}.

The existence of collective optic modes in molten salts has been widely predicted theoretically for more than four decades. However, in both the {\it ab initio} \cite{Bryk} and the classical \cite{Alcaraz} MD calculations, realistic IXS spectra were calculated for molten NaI by taking the respective weighting factors for the partial contributions into account. Although the $S_{\rm ZZ}(Q,\omega)$ partial spectrum comprises a prominent peak, the optic excitation is predicted to be just an extremely small contribution to the total $S(Q,\omega)$ of $3.7\times10^{-5}$ at $Q=1.75$ nm$^{-1}$ given in \cite{Alcaraz}. Thus, a high-intensity low-background IXS experiment is an essential  prerequisite to detect the LO excitation modes in molten salts. 

The present IXS and {\it ab initio} MD simulation results show lower excitation energies of the LO modes and a stronger damping as compared with the theoretical prediction from the classical MD simulations using rigid or polarizable models \cite{HansenBook}, where the electrons are frozen in the ionic cores or the ion polarization contributions are included into the effective potentials. A theoretical comparison is also given in \cite{Bryk}. The present results show that classical MD simulations cannot reproduce the present results, but only \textit{ab initio} MD simulations are capable of making 
predictions of the softening and damping of the LO modes in molten salts.

The origin of damping for the LO modes was discussed in a previous paper \cite{BrykJPCM}, where an analytical expression for $\Gamma$ is given. There, it is suggested that the main contribution comes from microscopic ionic conductivity. Since the ionic conduction in a molten salt occurs by releasing ions from solid-like cages formed instantaneously in the melt, the lifetime of the LO modes of $\sim0.3$ ps corresponds to the lifetime of the cages. 

In summary, we have presented the first reliable experimental data to prove the existence of LO collective modes in molten NaI in the low $Q$ range. The experiment was carried out using high-resolution IXS at BL35XU/SPring-8 where we were able to perform a high quality experiment with a low background. The obtained experimental results are in excellent agreement with theoretical predictions from an {\it ab initio} MD simulation. The observation of these modes at such small $Q$ and a good agreement with the simulation permit clear identification of these as collective optic modes with well defined phasing between different ionic motions. From the widths of the excitation peaks, the lifetimes of the LO modes were estimated to be about 0.3 ps, which would be related to the lifetime of the cages formed instantaneously around the respective ions.

\section*{Acknowledgements}
The authors thank Dr. F. Demmel of ISIS for giving motivation of this work and for helpful discussion. The authors thank Dr. A.Q.R. Baron for the
support in the experiment and helpful discussions.
This work was supported by Japan Society for the Promotion of Science (JSPS) Grant-in-Aid for Scientific Research (C) (No. 22540403) and by Japan Society for the Promotion of Science (JST) CREST (No.~JPMJCR1861). The IXS experiments were performed at the beamline BL35XU in the SPring-8 with the approval of the Japan Synchrotron Radiation Research Institute (JASRI) (No. 2009A1054).

\ukrainianpart

\title{Спостереження колективних збуджень оптичного типу в розплаві NaI}
\author{Ш. Хосокава\refaddr{label1,label2}, M. Iнуї\refaddr{label3}, T. Брик\refaddr{label4}, I. Mриглод\refaddr{label4}, В.-К. Пільгрім\refaddr{label2}, Ю. Kaджіхара\refaddr{label3}, K.~Maцуда\refaddr{label5}, Й. Омаса\refaddr{label6}, C. Цуцуї\refaddr{label7}}
\addresses{
\addr{label1} Фізичний факультет, Університет Кумамото, Кумамото 860-8555, Японія
\addr{label2} Хімічний факультет, Фізична хімія, Філіппс університет Марбурга, 35032 Марбург, Німеччина
\addr{label3} Школа інтегрованих мистецтв і наук, Університет Хіросіми, Хігаші-Хіросіма 739-8521, Японія
\addr{label4} Iнститут фізики конденсованих систем НАН України, 79011 Львів, Україна
\addr{label5} Фізичний факультет, Університет Кіото, Кіото 606-8502, Японія
\addr{label6} Лабораторія нейтронних досліджень, Інститут фізики твердого тіла, Університет Токіо, Kaшіва, Японія
\addr{label7} Японський дослідницький інститут синхротронної радіації (JASRI), Сайо 679-5198, Японія
}

\makeukrtitle

\begin{abstract}

Вимірювання непружного розсіювання рентгенівських променів високої роздільної здатності були
проведені на розплаві NaI біля точки плавлення при 680$^\circ$C в центрі SPring-8. Малі та згасаючі 
індикації повздовжних оптичних збуджень спостерігались в хвостах повздовжніх акустичних мод 
при малих імпульсах передачі, $Q\sim5$~нм$^{-1}$. Отримані спектри є в доброму узгодженні і по частоті,
і по ширині лінії з моделюванням методом першопринципної молекулярної динаміки (МД), а не з
класичними МД симуляціями. Спостереження цих мод при малих $Q$ та добре узгодження з моделюванням 
дозволяє чітку ідентифікацію їх як колективних оптичних мод з добре означеним фазуванням 
між рухами різних іонів.

\keywords  непружне розсіювання рентгенівських променів, розплав солі, оптичні збудження, першопринципне моделювання

\end{abstract}


\begin{thebibliography}{55}
\bibitem{Hansen} Hansen J.-P.,  McDonald I.R., Phys. Rev. A, 1975, {\bf 11}, 2111, \doi{10.1103/PhysRevA.11.2111}.
\bibitem{Adams} Adams E.M., McDonald I.R.,  Singer K., Proc. R. Soc. London, Ser. A, 1977, {\bf 357}, 37,\\ \doi{10.1098/rspa.1977.0154}.
\bibitem{Dixon} Dixon M., Philos. Mag. B, 1983, {\bf 47}, 531, \doi{10.1080/13642812.1983.11643261}.
\bibitem{Price} Price D.L., Copley J.R.D., Phys. Rev. A, 1975, {\bf 11}, 2124, \doi{10.1103/PhysRevA.11.2124}.
\bibitem{McGreevy} McGreevy R.L., Solid State Phys., 1987, {\bf 40}, 247, \doi{10.1016/S0081-1947(08)60693-1}.
\bibitem{HansenBook} Hansen J.-P., McDonald I.R.,  Theory of Simple Liquids, 3rd Ed., Academic Press, London, 2006.
\bibitem{BrykNaCl} Bryk T., Mryglod I., Phys. Rev. B, 2005, {\bf 71}, 132202, \doi{10.1103/PhysRevB.71.132202}.
\bibitem{Baron} Baron A.Q.R., Tanaka Y., Goto S., Takeshita K., Matsushita T.,  Ishikawa T., J. Phys. Chem. Solids, 2000, {\bf 61}, 461, \doi{10.1016/S0022-3697(99)00337-6}.
\bibitem{BaronInstrum} Baron A.Q.R., J. Spectrosc. Soc. Jpn., 2009, {\bf 58}, 205 (in Japanese), [Preprint \arxiv{0910.5764}, 2009 (in English)].
\bibitem{HosokawaPRL} Hosokawa S., Inui M., Kajihara Y., Matsuda K., Ichitsubo T., Pilgrim W.-C., Sinn H., Gonz\'{a}lez L.E., Gonz\'{a}lez~D.J., Tsutsui S.,  Baron A.Q.R., Phys. Rev. Lett., 2009, {\bf 102}, 105502,\\ \doi{10.1103/PhysRevLett.102.105502}.
\bibitem{11} Hosokawa S., Munejiri S., Inui M., Kajihara Y., Pilgrim W.-C.,  Ohmasa Y., Tsutsui S., Baron A.Q.R., Shimojo~F., Hoshino K., J. Phys.:  Condens. Matter, 2013, \textbf{25}, 112101, \doi{10.1088/0953-8984/25/11/112101}.
\bibitem{12} Hosokawa S., Inui M., Kajihara Y., Tsutsui S., Baron A.Q.R., J.  Phys.: Condens. Matter, 2015,  \textbf{27}, 194104,  \doi{10.1088/0953-8984/27/19/194104}.
\bibitem{Demmel} Demmel F., Hosokawa S., Pilgrim W.-C.,  Tsutsui S., Nucl. Instrum. Methods Phys. Res., Sect. B, 2005, {\bf 238}, 98, \doi{10.1016/j.nimb.2005.06.025}.
\bibitem{Tamura}  Tamura K., Inui M.,  Hosokawa S., Rev. Sci. Instrum., 1999, {\bf 70}, 144, \doi{10.1063/1.1149556}.
\bibitem{HosokawaCMP} Hosokawa S., Condens. Matter Phys., 2008, {\bf 11}, 71, \doi{10.5488/CMP.11.1.71}.
\bibitem{HosokawaRSI} Hosokawa S., Pilgrim W.-C., Rev. Sci. Instrum., 2001, {\bf 72}, 1721, \doi{10.1063/1.1338487}.
\bibitem{Bryk} Bryk T., Mryglod I., Phys. Rev. B, 2009, {\bf 79}, 184206, \doi{10.1103/PhysRevB.79.184206}.
\bibitem{DHO} F{\aa}k B., Dorner B., Physica B, 1997, {\bf 234--236}, 1107, \doi{10.1016/S0921-4526(97)00121-X}.
\bibitem{Alvarez} Alvarez  M., Bermejo F.J., Verkerk P.,  Roessli B., Phys. Rev. Lett., 1998, {\bf 80}, 2141,\\ \doi{10.1103/PhysRevLett.80.2141}.
\bibitem{Jahn} Jahn S., Suck J.-B., Phys. Rev. Lett., 2004, {\bf 92}, 185507, \doi{10.1103/PhysRevLett.92.185507}.
\bibitem{Bermejo} Bermejo F.J., Taylor J.W., McLain S.E., Bustinduy I., Turner J.F.C., Ruiz-Martin M.D., Cabrillo C.,  Fernandez-Perea R., Phys. Rev. Lett., 2006, {\bf 96}, 235501, \doi{10.1103/PhysRevLett.96.235501}.
\bibitem{Alcaraz} Alcaraz O., Trull\`{a}s J., J. Chem. Phys., 2010, {\bf 132}, 054503, \doi{10.1063/1.3298863}.
\bibitem{Sette} Sette F., Ruocco G., Krisch M., Masciovecchio C., Verbeni R.,  Bergmann U., Phys. Rev. Lett., 1996, {\bf 77}, 83, \doi{10.1103/PhysRevLett.77.83}.
\bibitem{BrykJPCM} Bryk T., Mryglod I., J. Phys.: Condens. Matter, 2004, {\bf 16}, L463, \doi{10.1088/0953-8984/16/41/L06}.
\end{thebibliography}
\end{document}